\documentclass[conference]{IEEEtran}
\IEEEoverridecommandlockouts
\usepackage{cite}
\usepackage{amsmath,amssymb,amsfonts}
\usepackage{algorithmic}
\usepackage{graphicx}
\usepackage{textcomp}
\usepackage{xcolor}

\usepackage{hyperref}
\usepackage{algorithm}
\usepackage{amssymb}
\usepackage[T1]{fontenc}
\usepackage{bm}
\usepackage[justification=centering]{caption}

\usepackage{verbatim}
\def\BibTeX{{\rm B\kern-.05em{\sc i\kern-.025em b}\kern-.08em
    T\kern-.1667em\lower.7ex\hbox{E}\kern-.125emX}}
\begin{document}

\title{Distributed Massive MIMO-Aided Task Offloading in Satellite-Terrestrial Integrated Multi-Tier VEC Networks\\
\author{Yixin~Liu,~Shaoling~Liang,~Kunlun Wang,~Wen~Chen,~Yonghui~Li,~and~George~K.~Karagiannidis}
\thanks{Y. Liu, and K. Wang are with the School of Communication and Electronic Engineering, East China Normal University, Shanghai 200241, China (e-mail: 51255904085@stu.ecnu.edu.cn; klwang@cee.ecnu.edu.cn). 

S. Liang is with the Guangxi Key Laboratory of Digital Infrastructure, Nanning 530000, China (e-mail: liangsl@gxi.gov.cn).

W. Chen is with the Department of Electronic Engineering, Shanghai Jiao Tong University, Shanghai 200240, China (e-mail: wenchen@sjtu.edu.cn).

Y. Li is with the School of Electrical and Information Engineering, The University of Sydney, Australia (e-mail: yonghui.li@sydney.edu.au). 

George K. Karagiannidis is with the Department of Electrical and Computer Engineering, Aristotle University of Thessaloniki, 54 124 Thessaloniki, Greece (e-mail: geokarag@auth.gr).}
}

\maketitle

\begin{abstract}
This paper proposes a distributed massive multiple-input multiple-output
(DM-MIMO) aided multi-tier vehicular edge computing (VEC) system. In particular, each vehicle terminal (VT) offloads its computational task to the roadside unit (RSU) by orthogonal frequency division multiple access (OFDMA), which can be computed locally at the RSU and offloaded to the central processing unit (CPU) via massive satellite access points (SAPs) for remote computation. By considering the partial task offloading model, we consider the joint optimization of the task offloading, subchannel allocation and precoding optimization to minimize the total cost in terms of total delay and energy consumption. To solve this non-convex problem, we transform the original problem into three sub-problems and use the alternate optimization algorithm to solve it. First, we transform the subcarrier allocation problem of discrete variables into convex optimization problem of continuous variables. Then, we use multiple quadratic transformations and Lagrange multiplier method to transform the non-convex subproblem of optimizing precoding vectors into a convex problem, while the task offloading subproblem is a convex problem. Given the subcarrier and the task allocation and precoding result, we finally find the joint optimized results by iterative optimization algorithm. Simulation results show that our proposed algorithm is superior to other benchmarks.
\end{abstract}

\section{Introduction}
Multi-tier computing is considered as one of
the promising solution for those applications and it leads the revolution of vehicular edge computing (VEC) to grow fast \cite{w}, \cite{tang}.
In the rapidly evolving era of vehicular communications, the Vehicle-to-Everything (V2X) paradigm has emerged as a cornerstone for the future of intelligent transportation systems\cite{8300313}. Despite the significant advancements in V2X technologies, the current state of vehicular networks is often characterized by fragmented connectivity and variable data rates, primarily due to the reliance on terrestrial infrastructure. The dynamic nature of vehicular traffic, with its high mobility and sporadic distribution, further complicates the provision of consistent and reliable communication services.

To overcome these limitations and enhance the reliability and scalability of VEC networks, satellite-terrestrial integration emerges as a viable and innovative solution\cite{3}. Satellite-terrestrial networks combine the ubiquitous coverage of satellites with the high-capacity and low-latency features of terrestrial networks, creating a synergistic infrastructure capable of supporting VEC applications\cite{4}. Moreover, the dynamic topology of low-Earth orbit (LEO) satellites and other satellite constellations with different orbits, offers reduced propagation delay and improved throughput compared to traditional geostationary systems, making them well-suited for real-time vehicular applications\cite{5}.
By leveraging satellite-terrestrial integration, VEC networks can offload tasks to satellite-linked edge servers, maintaining reliable communication and computation even in challenging environments. This integration also enhances network robustness and scalability, and massively distributed LEOs can deploy distributed massive multiple-input multiple-output (DM-MIMO) systems to further improve spectral efficiency and support a growing number of connected vehicles\cite{7}. 

Affected by the cell-free architecture in terrestrial networks, this work proposes a novel LEO cluster architecture based on DM-MIMO technology that allows terrestrial terminals to connect to satellite clusters\cite{9939157}. In a ground cell-free massive MIMO (CF-mMIMO) system, all base stations cooperate to communicate with the user in a cell-free manner. The technology can be applied to satellite networks based on LEO coverage, using ultra-dense deployment, high-speed inter-satellite links (ISLs) to achieve line-of-sight (LoS) connections to ground-based vehicle  terminals (VTs).

This paper merges the DM-MIMO architecture in LEO satellite networks and task offloading to enhance the energy-latency performance of VEC. The key contributions of this paper can be summarized as follows:
\begin{itemize}
\item We consider an air-ground computation and offloading architecture based on DM-MIMO assisted VEC. Due to the limited computation resource,  the task can be offloaded remotely to distributed satellite for computation or can be handled locally in the road side unit (RSU) to decrease the total computation consumption. The terrestrial communication relay on orthogonal frequency division multiple access (OFDMA) to improve spectrum utilization and anti-interference ability.
\item Then, the problem of weighted sum energy consumption and time delay minimization is proposed by jointly optimizing subchannel allocation, precoding design and task offloading.
\item To address the challenging non-convex optimization problem, it is decomposed into three sub-problems and tackled using an alternating optimization approach. Specifically, the RSU-satellite access point (SAP) association sub-problem is resolved through fractional programming, employing a quadratic transformation and the Lagrangian dual method, ultimately yielding an optimized closed-form solution. Moreover, the optimal computation task allocation sub-problem is addressed using a convex optimization method.
\end{itemize}

\section{System Model and Problem Formulation}
\subsection{System Model}
We consider a VEC system, which consists of a RSU, multiple VTs, and a SAP cluster. The SAP cluster is equipped with K SAPs and a central processing unit (CPU). The computational task from each VT can be handled locally in the RSU, or transferred to the CPU for remote computing with the help of K SAPs. Let $I=\{1,2…I\}$ be the set of all VTs and $K=\{1,2…K\}$ be the set of SAPs. For the convenience of calculation, we consider the number of antennas to be RSU is the same as the number in VTs, i.e., each RSU has $I$ antennas. Each VT and SAP is equipped with one antenna.

\subsection{Channel Model}
For the channel model of the links from VTs to the RSU and from the RSU to SAPs, we consider both of small-scale fading and large-scale fading.
\subsubsection{Small-Scale Fading}
the channel for the $i$th data stream from RSU to the $k$th SAP can be modeled as as\cite{8645336}:
\begin{small}
\begin{equation}
    h_{i,k}=\sqrt{L_{i,k}} (\sqrt{\frac{\kappa_{i,k}}{\kappa_{i,k}+1} } )h_{i,k}^{'}+\sqrt{\frac{1}{\kappa_{i,k}+1} } )h_{i,k}^{''},
\end{equation}
\end{small}
where $\kappa_{i,k}$ is the Rician K-factor, $h_{i,k}^{'}$ and $h_{i,k}^{''}$ are LoS and non-LoS (NLoS) components. The LoS component  $h_{i,k}^{'} = e^{j\phi _{i,k}} $, where $\phi _{i,k}\sim \mathcal{U}[-\pi,\pi]$ follows uniform distribution and represents the phase shift from the mobilities of the RSU and SAP.
We assume that the NLoS component follows Rayleigh distribution, i.e. $h_{i,k}^{''} \sim \mathcal{CN} (0,1)$. 
\subsubsection{Large-Scale Fading}
As for the large-scale fading for the $i$th data stream from RSU to the $k$th SAP, it can be expressed as follows:
\begin{small}
\begin{equation}
    L_{i,k}=10^{-(L_{i,k}^{\rm{dist}}[dB]+L_{i,k}^{\rm{shad}}[dB]+L_{i,k}^{\rm{ant}}[dB])/10}.
\end{equation}
\end{small}
\begin{itemize}
\item  $L_{i,k}^{\rm{dist}}[dB]$ is the path loss. It can be expressed as:
\begin{small}
\begin{equation}
    L_{i,k}^{\rm{dist}}[dB] = 32.45+20\log_{10}f_c+20\log_{10}d_{i,k},
\end{equation}
\end{small}
where $f_c$ and $d_{i,k}$ respectively represent the carrier frequency and the distance between $i$th antenna of the RSU and SAP $k$.
\item $L_{i,k}^{\rm{shad}}[dB]$ is the shadow losses. We model it as a long-normal variable (RV), i.e. $L_{i,k}^{\rm{shad}}[dB]\sim \mathcal{N}(0,\sigma_{sh}^2)$, where $\sigma_{sh}^2$ is the shadowing variance.
\item $L_{i,k}^{\rm{ant}}[dB]$ is the antenna losses due to the boresight angle and can be expressed as\cite{8966367}
\begin{small}
\begin{equation}
    L_{i,k}^{\mathrm{ant}}=-10\log_{10}(\cos(\theta_{i,k})^\eta\frac{32\log2}{2({2\arccos(\sqrt[\eta]{0.5}} ))^2}).
\end{equation}
\end{small}
\end{itemize}
\subsection{Problem Formulation}
In our proposed VEC system, due to the limited computational resources of VTs, all tasks from VTs are offloaded to RSU by OFDMA technique, and task computation consists of two parts: one part is left in RSU for local computation; another part of the task is transferred from RSU to SAPs through DM-MIMO technique, and finally converged to the CPU for computation.
\subsubsection{VTs to RSU}
In the terrestrial scenario, we can offload the task to the RSU by OFDMA. 
Due to the intractability of the discrete variables, we transform the 0-1 variables of subchannel allocation into continuous variables within the interval $[0,1]$ based on \cite{8438896}. So the transmission rate between VT $i$ and RSU can be stated as:
\begin{small}
\begin{equation}
    R_i^{\mathrm{R}}=\alpha_iB{\log}_2(1+\frac{g_i^{\mathrm{R}}p_i^{\mathrm{R}}d_i^{-\varepsilon}}{{\sigma_n^{\mathrm{R}}}^2}),
\end{equation}
\end{small}
where $B$ represents the total bandwidth, and $\alpha_i$ represents the variable used to allocate different subcarrier bandwidth to each VT. $g_i^R$ is the channel gain between VT $i$ and RSU. We use $d_i^{-\varepsilon}$ to denote the mobility of VT, where $\varepsilon$ represents the path loss in the terrestrial layer, the $d_i$ denotes the distance between VT $i$ and the RSU.

Then the transmission delay from $i$th VT to the RSU can be defined as:
\begin{small}

\begin{equation}
    t_{i,\mathrm{trans}}^{\mathrm{R} }=\frac{S_{i}^{\mathrm{tot} }}{R_i^{\mathrm{R}} }. 
\end{equation}   
\end{small}
In addition, another part of the computation task for the $i$th VT processed in the RSU requires a computation time of
\begin{small}
\begin{equation}
    t_{i,\mathrm{cp}}^{\mathrm{R} }=\frac{S_{i}^{\mathrm{R} }}{f_i^{\mathrm{R}} }. 
\end{equation}
\end{small}
For the local computation, the energy consumption includes the transmit energy consumption from the VT to RSU, and the computation energy consumption in the RSU, which is given by:
\begin{small}
\begin{align}
    E^{\mathrm{R}} &=E_{\mathrm{cp}}^{\mathrm{R}}+E_{\mathrm{trans}}^{\mathrm{VT}}\\\notag
    &=\sum_{i\in I}{p_i^{\rm{R}}\cdot\frac{S_i^{\mathrm{tot} }}{R_i^{\mathrm{R}} }+\rho_{\mathrm{R}}\cdot{f^{\mathrm{R}}}^2\sum_{i\in I} S_i^\mathrm{R}},
\end{align}
\end{small}
where $p_i^{\rm{R}}$ represents the transmission power of VT $i$ to RSU, $S_{i}^{\mathrm{R} }$ is the task allocated to RSU, $f_i^{\mathrm{R}} $ is the RSU computational frequency and $\rho_{\mathrm{R}}$ is the effective switching capacitance coefficient of the RSU.
\subsubsection{RSU to SAPs}
In a SAP cluster, the SAP is connected with the CPU by ISLs. The ISLs is very high-speed, so we ignore the delay and the energy cost of the duration. As for the remote task computation, we assume all the tasks will converge to the CPU for computation.

In the process of transmitting tasks to the SAP, the transmission rate of task offloading is:
\begin{small}
\begin{equation}
    R_i^{\rm{S} }= B_s \log_2(1+\frac{\left|E\{\mathbf{h}_\mathit{i} \mathbf{v}_\mathit{i} \}\right|^2}{ {\textstyle \sum_{\mathit{j} \in I\setminus \mathit{i} }\left|E\{\mathbf{h}_\mathit{j} \mathbf{v}_\mathit{j} \}\right|+{\sigma _{n}^{\rm S}} }^2} ),
\end{equation}
\end{small}
where $ B_s $ represents the channel bandwidth for the transmission process, $h_i=[h_{i,1},h_{i,2},....]$ denotes the channel coefficient, $ \mathbf{v}_i$ and $\sigma_{n}^{{\rm S}} $ respectively denote the precoding vector and the noise in the transmission channel. 

So the remote computation delay includes the transmission delay from RSU to SAPs and the computation delay at the CPU, which is given by:
\begin{small}
\begin{equation}
      t_i^{\mathrm{S} }=\frac{S_{i}^{\mathrm{S} }}{R_i^{\mathrm{S}} } +\frac{{S_{i}^{\rm S}}}{f^{\mathrm{CPU}} },
\end{equation}
\end{small}
where $S_{i}^{\mathrm{S} }$ and $f^{CPU}$ respectively represent the task allocated to SAPs and the CPU computational frequency.
Similarly, the remote computation energy consumption consists of the transmission energy consumption and the CPU computation energy consumption:
\begin{small}
\begin{align}
    E^{\mathrm{S}} =E_{\mathrm{cp}}^{\mathrm{S}}+E_{\mathrm{trans}}^{\mathrm{S}} =\rho_{\mathrm{S}}\cdot{f^{\mathrm{CPU}}}^2\sum_{i\in I} S_i^\mathrm{S} +p_i^S \cdot\frac{\sum_{i \in I}{S_i^{\mathrm{S}}}}{R_i^{\mathrm{S}}},
\end{align}
\end{small}
where $\rho_{\mathrm{S}}$ represents the effective switching capacitance coefficient of the CPU in the SAPs' cluster. $f^{\mathrm{tot}} $ is the RSU computational frequency, $p_i^{\rm{S}}$ represents the transmission power of RSU to SAPs. Specifically, we set $p_i^{\rm{S}} = {\left \| \mathbf{v}_i  \right \| } ^2_F$. Since the computation result is very small, we ignore the receiving cost.

The total delay and energy consumption are given by:
\begin{small}
\begin{equation}
    t_i^{\rm{tot}}=t_{i}^{\rm{trans}}+\max\{t_i^{\rm{S}},t_{i,\mathrm{cp}}^{\rm{R}}\}\label{t},
\end{equation}
\begin{equation}
    E^{\rm{tot}} = E^{\mathrm{R}}+E^{\mathrm{S}}.
\end{equation}
\end{small}

The task offloading decision variables, precoding vector variables, and the subchannel bandwidth allocation
variables can be denoted as $\mathbf{S}=\{S_i^{\mathrm{R} },S_i^{\mathrm{S}  }\}_{\forall i \in I}$, $\mathbf{V}={\{ \mathbf{v} _i\}}_{\forall i \in I }$ and $\bm{\alpha} ={\{\alpha } _i\}_{\forall i \in I }$, respectively. Let
$\beta \in [0, 1]$ denote the weight factor between delay and
energy consumption, which affects the weight of them in the total cost of the system. Then, the
total cost can be denoted as:
\begin{small}
\begin{subequations}\label{total}
\begin{align}
\min_{\mathbf{S,V,\alpha}}\ &d_\Omega =\beta t_i^{\rm tot}+(1-\beta)E^{\rm tot}\label{1}\\
\rm {s.t.}\,\,
&S_{i}^{\rm{R}}+S_{i}^{\rm{S}} = S_i^{\rm tot},S_{i}^{\rm{R}}\ge 0,S_{i}^{\rm{S}}\ge 0,\label{2}\\
&{\left \| \mathbf{v}_i  \right \| } ^2_F \le P^{\rm{max}},\label{3}\\    &\sum_{i\in \mathcal{I} }\alpha_{i}\le 1,\label{4}\\
 &t_i^{\mathrm{tot}}\le T^{\mathrm{max}}\label{5}.
\end{align}
\end{subequations}
\end{small}
Constraint \eqref{2} ensures that the total amount of computational
task offloaded by the task.
Constraint \eqref{3} ensures that the norm's square of precoding vector cannot exceed the maximum power $P^{\rm{max}}$.
Constraint \eqref{4} ensures that the sum of allocation coefficient
cannot exceed 1.
Constraint \eqref{5} gives the constraint to the total time delay and it cannot exceed the maximum tolerable delay $T^{\rm{max}}$.

\section{Proposed Optimization Solution}
In this section, the optimization problem is analyzed and solved. We decouple the intractable non-convex Problem \eqref{total} into three subproblems, which are then iteratively optimized.

\subsection{Optimizing subchannel allocation While Fixing the Rest Optimization Variables}
We first solve the subchannel allocation subproblem with a given precoding vector $\mathbf{v}_i$. To simplify \eqref{1}, we define $A_i$ and $t_c$ as:
\begin{small}
\begin{subequations}
    \begin{align}
    A_i&= E_{\mathrm{cp}}^{\mathrm{R}}+E_{\mathrm{cp}}^{\mathrm{S}}+E_{\mathrm{trans}}^{\mathrm{S}},\\
    t_c&= \max\{t_i^{\rm{S}},t_{i,\mathrm{cp}}^{\rm{R}}\}.
   \end{align}
\end{subequations}
\end{small}
Then we relax the discrete variables of subcarrier allocation into continuous variables, the subchannel allocation subproblem can be transformed into:
\begin{small}
\begin{subequations}\label{eqn-1}
\begin{align}
    \min_{\mathbf{\alpha}} d_\Omega =\beta &(\frac{S_{i}^{\mathrm{tot} }}{\alpha_iB{log}_2(1+\frac{g_i^{\mathrm{R}}p_i^{\mathrm{R}}d_i^{-\varepsilon}}{{\sigma_n^{\mathrm{R}}}^2})}+ t_c) \\\notag
    +(1-\beta)(A_i+&\sum_{i\in I}{p_i^{\rm{R}}\cdot\frac{S_i^{\mathrm{R} }}{\alpha_i B{log}_2(1+\frac{g_i^{\mathrm{R}}p_i^{\mathrm{R}}d_i^{-\varepsilon}}{{\sigma_n^{\mathrm{R}}}^2}) }})\\
    \rm {s.t.}\,\,
    &\sum_{i\in \mathcal{I} }\alpha_{i}\le 1,\\
    &t_i^{\mathrm{tot}}\le T^{\mathrm{max}}.
\end{align}
\end{subequations}
\end{small}
The above Problem \eqref{eqn-1} is a convex optimization problem, which can be easily solved\cite{cvx}.

\subsection{Optimizing precoding vector While Fixing the Rest Optimization Variables}
After solving the subcarrier allocation problem, we substitute the obtained suboptimal solution back into the original optimization problem. To simplify \eqref{1}, we define $C_i$ as:
\begin{small}
\begin{equation}
    C_i= E_{\mathrm{cp}}^{\mathrm{R}}+E_{\mathrm{cp}}^{\mathrm{S}}+E_{\mathrm{trans}}^{\mathrm{VT}}.
\end{equation}
\end{small}
To assist optimizing this subproblem, an auxiliary variable $t=t_i^{\rm{tot}}$ is introduced. So the precoding vector subproblem can be obtained as:
\begin{small}
\begin{subequations}\label{eqn-2}
\begin{align}
    \min_{\mathbf{\mathbf{v}_k,t}} d_{\Omega_2} =&\beta t\\\notag
    +(1-\beta)&(C_i+\sum_{i\in I}{p_i^{\rm{R}}\cdot\frac{S_i^{\mathrm{S} }}{B_s \log_2(1+\frac{\left|E\{\mathbf{h}_\mathit{i} \mathbf{v}_\mathit{i} \}\right|^2}{ {\textstyle \sum_{\mathit{j} \in I\setminus \mathit{i} }\left|E\{\mathbf{h}_\mathit{i} \mathbf{v}_\mathit{i} \}\right|+{\sigma _{n}^S} }^2}})}\\
    \rm {s.t.}\,\,\,
    &\frac{S_{i}^{\mathrm{S} }}{B_s \log_2(1+\frac{\left|E\{\mathbf{h}_\mathit{i} \mathbf{v}_\mathit{i} \}\right|^2}{ {\textstyle \sum_{\mathit{j} \in I\setminus \mathit{i} }\left|E\{\mathbf{h}_\mathit{i} \mathbf{v}_\mathit{i} \}\right|+{\sigma _{n}^S} }^2} ) } +\frac{S_{i}^{\mathrm{S} }}{f_i^{\mathrm{S}} }\le t,\\
    &\sum_{i\in \mathcal{I} }\alpha_{i}\le 1 \label{22c},\\
    &t\le T^{\mathrm{max}}\label{22d}.
\end{align}
\end{subequations}
\end{small}
It is obvious that Problem (18) is still non-convex. Then, we use the quadratic transform \cite{8314727} to solve problem \eqref{eqn-2}. By introducing the auxiliary variable $y_i$, problem \eqref{eqn-2} can be transformed into:
\begin{small}
\begin{align}\label{eq-3}
     \min_{\mathbf{\mathbf{v}_i,t,y_i}} d_{\Omega_2} = \beta t+2(1-\beta)y_i\sqrt{S_i^{\rm{S}}\cdot{\left \| \mathbf{v}_i  \right \|}_F^2 } \\\notag
     -(1-\beta)y_i^2R_i^{\rm{S}}(\mathbf{v}_i)+(1-\beta)C_i\\\notag
     \rm {s.t.}\,\,\, (22b)-(22d).
\end{align}
\end{small}
However, problem \eqref{eq-3} is still intractable. It is mainly due to the fractional item of $R_i^{\rm{S}}(\mathbf{v}_i)$. We observe that when Problem \eqref{eq-3} meets its minimum, the $R_i^{\rm{S}}(\mathbf{v}_i)$ meets its maximum. As a result, we proposed to change $R_i^{\rm{S}}$ into another trackable form. We let $R_i^{\rm{S}}(\mathbf{v}_i) =\log_2{ (1+{\frac{M_i(\mathbf{v}_i)}{N_i(\mathbf{v}_i)}}}) $, then we introduce  $\gamma_i $  to replace the ratio term inside,  so the new form of $R_i^{\rm{S}}(\mathbf{v}_i)$ is as follows \cite{8314727}:
\begin{small}
\begin{subequations}\label{24}
\begin{align}
    &\max_{\gamma_i}  B_s\log_2(1+\gamma_i)\\
    &{\rm{s.t.}}\ \ \gamma_{i} \le \frac{M_i(\mathbf{v}_i)}{N_i(\mathbf{v}_i)} .\label{24b}
\end{align}
\end{subequations}
\end{small}
Obviously, the solution for this inner optimization problem is that $\gamma_i$ should satisfy \eqref{24b} with equality. Note the \eqref{24} has the characteristic of strong duality because of the convexity in $\gamma_i$, we introduce the dual variable $\lambda_i$ and formulate the Lagrangian function:
\begin{small}
\begin{equation}\label{25}
    L(\gamma_i,\lambda _i)=B_s\log_2(1+r_i)-\lambda_i(\gamma_i-\frac{M_i(\mathbf{v}_i)}{N_i(\mathbf{v}_i)} )
\end{equation}
\end{small}
Due to the strong duality, Problem \eqref{25} is equivalent to the following dual problem:
\begin{small}
\begin{equation}\label{26}
    \min_{\lambda _i\ge 0} \max_{\gamma_i} L(\gamma_i,\lambda _i).
\end{equation}
\end{small}
We assume $(\gamma_i^*, \lambda_i^*)$ is the saddle point of the problem \eqref{26}. So it has to satisfy the first derivative condition $\frac{\partial L}{\partial \gamma_i} =0$, we have:
\begin{small}
\begin{equation}
     \lambda_i^*=\frac{B_{\rm s}}{1+\gamma_i^*}, \forall i = 1,...,I.
\end{equation}
\end{small}

Through the solution of optimization problem \eqref{24}, we already have the result $\gamma_i=\frac{M_i(\mathbf{v}_i)}{N_i(\mathbf{v}_i)}$:
\begin{small}
\begin{equation}\label{28}
    \lambda_i^{*}=\frac{B_{\rm{s}}N_i(\mathbf{v}_i)}{M_i(\mathbf{v}_i)+N_i(\mathbf{v}_i)},\forall i= 1,...,I
\end{equation}
\end{small}
the reformulated objective function $\tilde{{R_i^{\mathrm{S} }} }(\mathbf{v}_i,\gamma_i )$ can be defined as:
\begin{small}
\begin{equation}\label{29}
    \tilde{{R_i^{\mathrm{S} }} }(\mathbf{v}_i,\gamma_i )= B_{\rm{s}}\log_2(1+\gamma_i)+\frac{B_{\rm{s}}M_i(\mathbf{v}_i)(1+\gamma_i)}{M_i(\mathbf{v}_i)+N_i(\mathbf{v}_i)} -B_{\rm{s}} \gamma_i.
\end{equation}
\end{small}
The expressions \eqref{29} and \eqref{24} are equivalent. Observe that  $\tilde{{R_i^{\mathrm{S} }} }(\mathbf{v}_i,\gamma_i )$ is a convex function over $\gamma_i$ when the $\mathbf{v}_i$ is fixed. We substitute the $\gamma_i^*=\frac{M_i(\mathbf{v}_i)}{{N_i(\mathbf{v}_i)}} $ into $\tilde{{R_i^{\mathrm{S} }} }(\mathbf{v}_i,\gamma_i )$, the equivalence is therefore established.

Based on the above transforms, we still use the quadratic transform to further recast $ \tilde{{R_i^{\mathrm{S} }} }(\mathbf{v}_i,\gamma_i )$ to $ \tilde{{R_i^{\mathrm{S} }} }(\mathbf{v}_i,\gamma_i ,z_i)$ as:
\begin{small}
\begin{align}
    \tilde{{R_i^{\mathrm{S} }} }(\mathbf{v}_i,\gamma_i ,z_i)=  &2z_i\sqrt{B_{\rm{s}}M_i(\mathbf{v}_i )(1+\gamma_i)}-z_i^2[M_i(\mathbf{v}_i )+N_i(\mathbf{v}_i)]\\\notag
    &+B_{\rm{s}}\log_2(1+\gamma_i)-B_{\rm{s}}\gamma_i.
\end{align}
\end{small}
It can be observed directly that $\sqrt{B_{\rm{s}}M_i(\mathbf{v}_i )(1+\gamma_i)}$ and $ B_{\rm{s}}\log_2(1+\gamma_i)-B_{\rm{s}}\gamma_i$ are both convex with respect to $\gamma_i$. In addition, $z_i^2[M_i(\mathbf{v}_i )+N_i(\mathbf{v}_i)]$ is quadratic with respect to $z_i$ and $\mathbf{v}_i $ separately. Therefore, it is convex over both. Hence, the function $ \tilde{{R_i^{\mathrm{S} }}}(\mathbf{v}_i,\gamma_i,z_i)$ exhibits convexity with respect to the variables $\mathbf{v}_i$, $\gamma_i$, and $z_i$, each considered independently.

The problem \eqref{eqn-2} is reformulated as:
\begin{small}
\begin{subequations}\label{eqn-3}
    \begin{align}
        &\min_{\mathbf{v} _i,t,y_i,\gamma_i,z_i}  d_{\Omega_2}= \\\notag
       & (1-\beta) \sum_{i\in I}[2y_i S_i^{\rm{S}}\left \| \mathbf{v} _i \right \|_F^2 -y_i^2 \tilde{{R_i^{\mathrm{S} }} }(\mathbf{v}_i,\gamma_i ,z_i)] +\beta t +(1-\beta)C_i\\
       &\rm{s.t.}\ \ \
       \frac{S_{i}^{\rm{S} }}{\tilde{{R_i^{\rm{S} }} }(\mathbf{v}_i,\gamma_i ,z_i)} +\frac{S_{i}^{\rm{S} }}{\mathit{f}_i^{\rm{S}} }\le t, \\ \notag
       &\ \ \ \ \ \ \ \ \ (22c),(22d).
    \end{align}
\end{subequations}
\end{small}
Due to the fact that $\tilde{{R_i^{\rm{S} }} }(\mathbf{v}_i,\gamma_i ,z_i)$ is a convex function with respect to the variables $z_i$ and $\mathbf{v}_i $ separately. Problem \eqref{eqn-3} is a convex problem, so the optimum $\mathbf{v}_i$ can be found through CVX. Thus we iteratively optimize problem \eqref{eqn-3}. When other variables are fixed, the optimized $z_i$ and $y_i$ are given by:
\begin{small}
\begin{align}
    z_i^* =  \frac{\sqrt{B_{\rm{s}}M_i(\mathbf{v}_i )(1+\gamma_i)}}{M_i(\mathbf{v}_i )+N_i(\mathbf{v}_i)}, \\
    y_i^* =  \frac{\sqrt{S_i^{\rm{S}}\left \| \mathbf{v} _i \right \|_F^2  }}{ \tilde{{R_i^{\mathrm{S} }} }.(\mathbf{v}_i,\gamma_i ,z_i)}. 
\end{align}
\end{small}
Given the obtained solution of $\gamma_i,z_i$ and $y_i$, the optimized $\mathbf{v} _i$ can be obtained by solving Problem \eqref{eqn-3}.
\subsection{Alternate optimization algorithm}
Note that the task allocation subproblem is a convex problem, we can use CVX to solve it. So the total iterative optimization algorithm can be summarized in Algorithm 1.

\begin{algorithm} 
	\caption{Alternate Optimization} 
        \renewcommand{\algorithmicrequire}{\textbf{Input:}}
	\renewcommand{\algorithmicensure}{\textbf{Output:}} 
	\begin{algorithmic}
		\REQUIRE Initialize ${S_i^{\rm S}}^{(0)}, {\alpha_i}^{(0)}$ and ${v_i}^{(0)}$ feasible values; iteration number: n=1; tolerance $\epsilon$.
		\ENSURE Optimized solutions: \bm{${{S}_i^{\rm S}}^\ast$},\bm{${\alpha_i}^\ast$}, \bm{${v_i}^\ast$};
            \WHILE{ the objection value ${f_{qm}}^{(n)}-{f_{qm}}^{(n-1)} \le \epsilon$ or $n\le 1000$}\label{whilestart}
		\STATE1.  Obtaining the optimal \bm{${\alpha_i}^{(n)}$} by solving the problem \eqref{eqn-1} under given \bm{${{S}_i^{\rm S}}^{(n-1)}$} and \bm{${v_i}^{(n-1)}$}.
        
		\STATE2. Obtaining the optimal \bm{${v_i}^{(n)}$} by solving the problem \eqref{eqn-3} under given \bm{${{S}_i^{\rm S}}^{(n-1)}$} and \bm{${\alpha_i}^{(n-1)}$}.
            \STATE3. Obtaining the optimal \bm{${S_i^{\rm S}}^{(n)}$} by CVX  under given \bm{${\alpha_i}^{(n-1)}$}, \bm{${v_i}^{(n-1)}$} .\\ $n=n+1$
            \ENDWHILE
	\end{algorithmic}
\end{algorithm}

\section{simulation results}
In this section, we quantify the accuracy of the analytical results by means of simulations. We set the satellites' altitude to 550km, antenna factor $\eta$ to be 10 wavelengths and Rician K-factor to be 10dB \cite{8966367}. The shadowing std is set to be 5dB and the carrier frequency is 12GHz.
The background noise density is configured at -174dBm/Hz. The input data size for each VT is uniformly distributed between 0.8MB to 1.2MB. E Each VT transmits with a power of $p_i^{\rm R} = 0.1$W, while the RSU operates with a maximum transmit power of $P^{\rm max} = 2$W. The RSU's computational capability is set to 0.8GHz and the SAP is set as 20GHz. The bandwidth are $B=2$MHz and $B_s=200$MHz. Unless stated otherwise, the other parameters are set as follows: $I=4,K=8,\beta =0.7$, and $\rho_R = \rho_S = 10^{-25}.$
For comparison, we consider the three other offloading methods: 1) the RSU-only scheme, where all the tasks are computed locally in the RSU. 2) SAPs-only, where all the tasks are transmitted remotely to the CPU via SAPs. 3) random allocation, where the tasks are offloaded randomly.

Fig. 1 illustrates the system cost, expressed as a weighted total delay and energy consumption, versus the number of iterations. This analysis validates the algorithm's convergence under varying numbers of VTs and different maximum transmission power between the RSU and SAPs. As shown in Fig. 1, the proposed algorithm converges within approximately four iterations and the weighted total cost increases with VT number $I$. Notably, for the same number of VTs, the convergence iteration count remains almost consistent across different maximum transmission power settings, demonstrating that the algorithm effectively minimizes the total cost.

\begin{figure}[htbp]
    \centering
    \includegraphics[width=2.6 in]{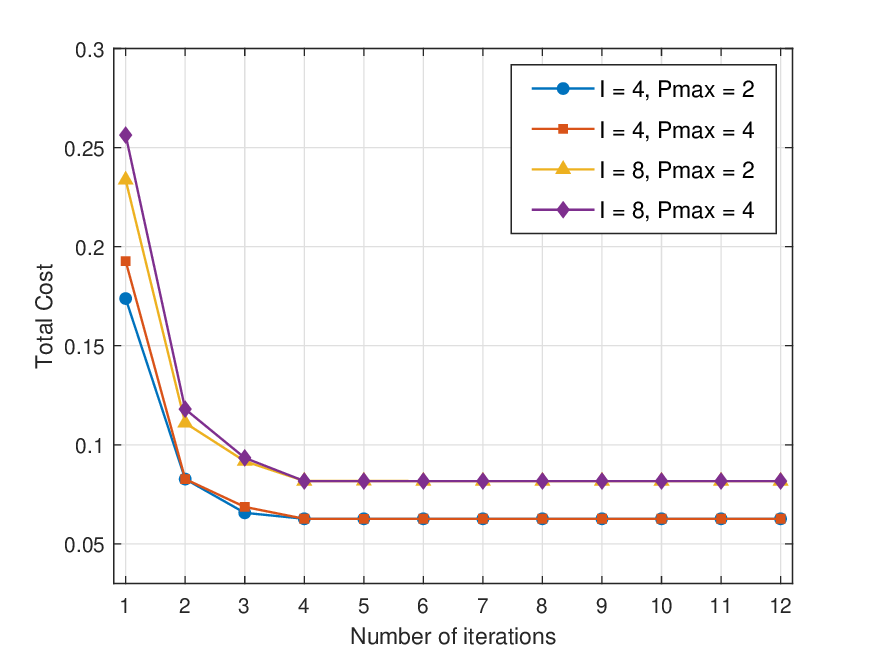}
    \caption{Convergence curve.}
    \label{fig:enter-label}
\end{figure}

\begin{figure}
	\begin{minipage}[t]{0.49\linewidth}
		\centering
		\includegraphics[width=1.85in]{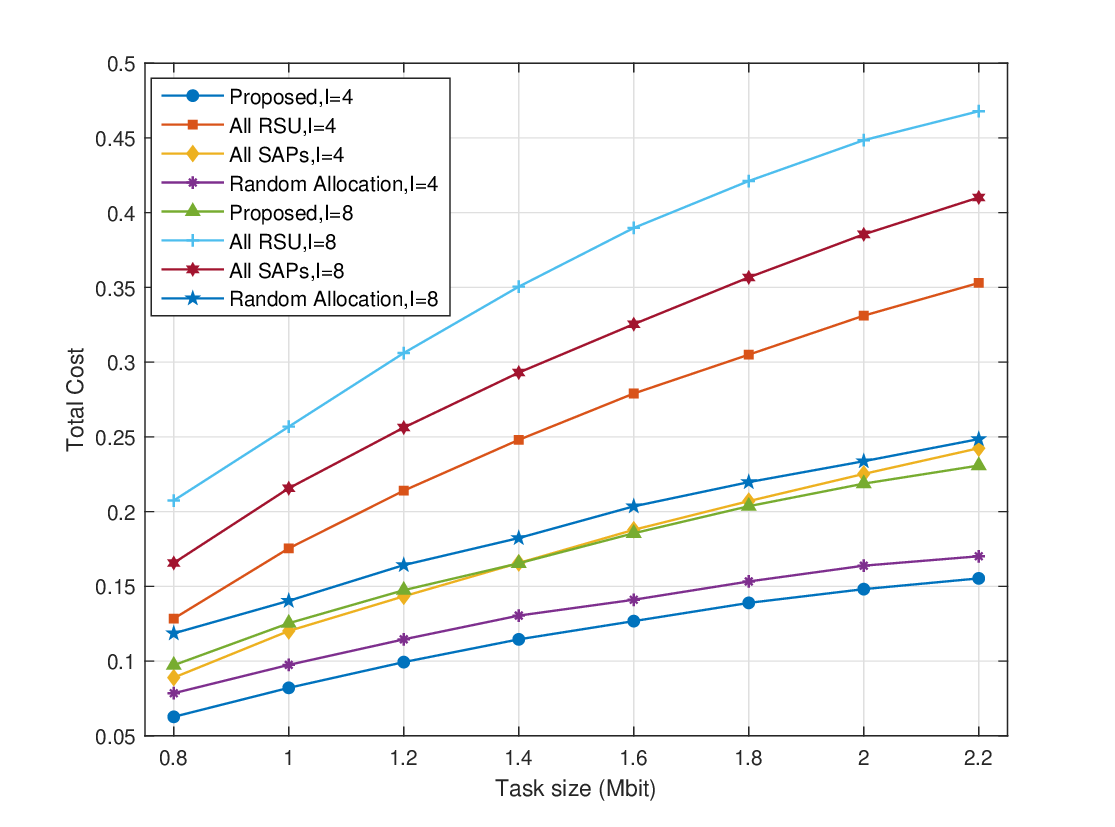}
            \captionsetup{font={scriptsize}}
		\caption{Performance comparison under different task data size.}
		\label{fig:hor_2figs_2cap_1}
	\end{minipage}
	\begin{minipage}[t]{0.49\linewidth}
		\centering
		\includegraphics[width=1.75in]{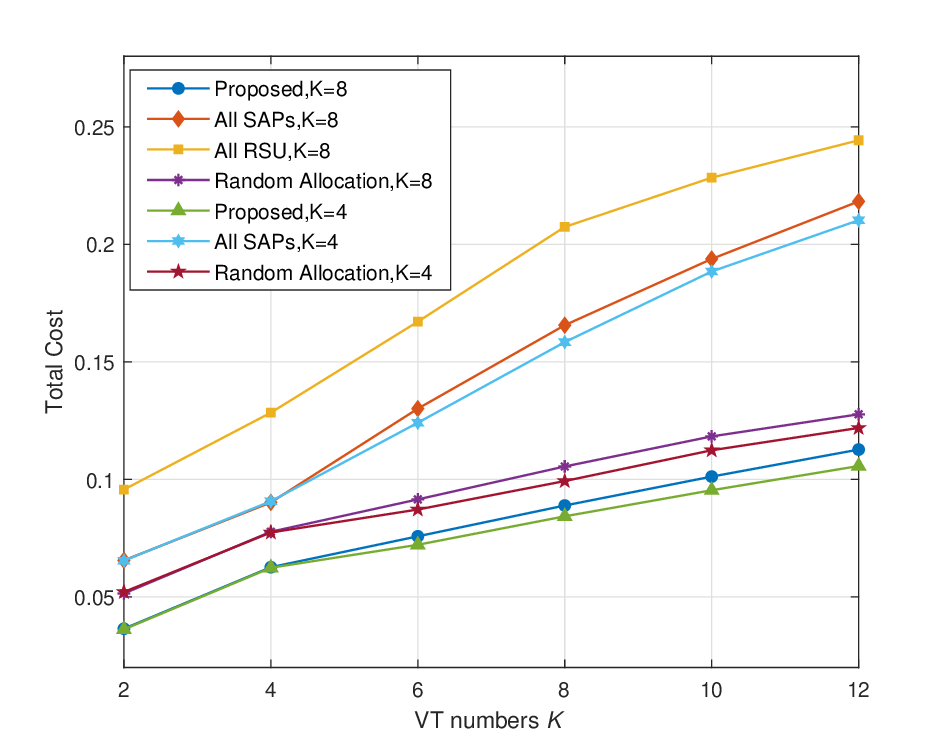}
            \captionsetup{font={scriptsize}}
		\caption{Performance comparison under different VT numbers.}
		\label{fig:hor_2figs_2cap_2}
	\end{minipage}
\end{figure}
As shown in Fig. 2, we can observe that when the VT number is the same, the performance of the proposed algorithm is better than that of the benchmark algorithms as the task scale increases. Owing to RSU's restricted computing capacity, the overall cost of local RSU computing rises more quickly in proportion to task size and surpasses that of the remote SAPs-only scheme. Excluding these extreme offloading scenarios, we also considered the scenario of random task offloading. We can observe that when the offloading ratio is random, the total cost is larger than that of the proposed algorithm.
It can be observed that as the number of VTs increases, the total cost for the three different offloading schemes also increases, while the performance advantage of our proposed algorithm over the benchmark algorithms remains the same.

In Fig. 3, it is apparent that as the number of VTs increases, the total cost also increases. This outcome is expected, as an increase with the amount of computational tasks leads to larger energy consumption and total delay. Additionally, for a fixed number of VTs, the total cost will rise slightly as the number of SAPs increases. Therefore, when the number of vehicles is fixed, the number of SAP has little impact on energy consumption, and we can reduce the total cost of the VEC system by increasing the number of SAP.

\section{conclusion}

In this paper, we propose a two-layer satellite-terrestrial distributed massive multiple-input multiple-output (DM-MIMO) assisted multi-tier vehicular edge computing (VEC) system. To optimize the system's performance, we formulate a problem to minimize the weighted sum of energy consumption and total delay, considering task offloading, subchannel allocation and precoding optimization across both the terrestrial and satellite tiers. This challenging optimization problem is tackled using an iterative solution approach, leveraging quadratic transformation and Lagrangian dual methods for effective problem reformulation and solution derivation. Simulation results demonstrate that our proposed architecture effectively meets the computing demands of vehicular terminals with low cost, while efficiently utilizing resources within multi-tier computing scenarios.



\footnotesize
\bibliographystyle{IEEEtran} 
\bibliography{IEEEabrv,ref} 

\end{document}